\documentclass[12pt]{article}
\usepackage{amsmath, amssymb,amsthm}

\newtheorem{Theorem}{Theorem}[section]
\newtheorem{Definition}[Theorem]{Definition}
\newtheorem{Remark}[Theorem]{Remark}
\newtheorem{Proposition}[Theorem]{Proposition}

\numberwithin{equation}{section}

\begin{document}

\centerline{\Large \bf A Galerkin approximation scheme for the mean correction}
\smallskip
\centerline{\Large \bf  in a mean-reversion stochastic differential equation}
\bigskip
\bigskip
\centerline{\bf Jiang-Lun Wu$^{a}$ and Wei Yang$^b$}
\smallskip
\centerline{${}^a\,$ Department of Mathematics, Swansea University }
\centerline{Singleton Park, Swansea SA2 8PP, UK ({\it j.l.wu@swansea.ac.uk})}
\smallskip
\centerline{${}^b\,$ Department of Mathematics and Statistics, University of Strathclyde }
\centerline{Glasgow G1 1XH, UK  ({\it w.yang@strath.ac.uk})}

\bigskip
\begin{abstract}
This paper is concerned with the following Markovian stochastic differential equation of mean-reversion type
\[
	dR_t= (\theta +\sigma \alpha(R_t, t))R_t dt +\sigma R_t dB_t
\]
with an initial value $R_0=r_0\in\mathbb{R}$, where $\theta\in\mathbb{R}$ and $\sigma>0$ are constants, and the mean correction function $\alpha:\mathbb{R}\times[0,\infty)\mapsto\alpha(x,t)\in\mathbb{R}$ is twice continuously differentiable in $x$ and continuously differentiable in $t$.  We first derive that under the assumption of path independence of the density process of Girsanov transformation for the above stochastic differential equation, the mean correction function $\alpha$ satisfies a non-linear partial differential equation which is known as the viscous Burgers equation. We then develop a Galerkin type approximation scheme for the function $\alpha$ by utilizing truncation of discretised Fourier transformation to the viscous Burgers equation. 

\end{abstract}

{\medskip\par\noindent
{\bf Mathematics Subject Classification (2000)}: 60H35; 35Q53

\smallskip\par\noindent
{\bf Key Words and Phrases}:  Markovian stochastic differential equation of mean-reversion type, 
viscous Burgers Equation,  truncation of (discretised) Fourier transformation, numerical approximation 
scheme.}

\section{Introduction}

Stochastic differential equations (SDEs) have become more and more popular in mathematical modeling the (random) 
dynamics involving uncertainty with the passage of time. There are diverse applications ranging from physics, biology, medical 
and health studies, climate studies, engineering, systematical science to economics and finance (cf. e.g. \cite{Oksendal} and 
references therein). In many studies of such stochastic models, the mean-reversion type stochastic dynamics is very important,  
which corresponds to equilibrium state(s) of the systems concerned and usually links to certain partial differential equations 
appeared in (classical) mathematical physics. 

In this paper, we study a stochastic differential equation of mean-reversion type which arises from the (modern) mathematical 
modeling in economics and finance. We aim to derive an equation for the mean correction function appearing in the SDE which characterizes the path inpendent property of the density process of the Girsanov transformation for the 
stochastic equation. Applying It\^o stochastic calculus, we end up with a viscous Burgers equation for the mean correction function.   
We then develop a Galerkin type approximation scheme for the mean correction function by utilizing Galerkin truncation of the 
(discrete) Fourier transform of the viscous Burgers equation. Our approximation provides an adaptive algorithm towards 
numerical solutions for the mean correction function and hence gives a way (as one may hope) to explore statistical behaviors 
of the mean-reversion type SDEs in the financial modeling.    
  
The rest of the paper is organized as follows. Section 2 will introduce the stochastic differential equations of mean-reversion 
type along with Girsanov transformation. We prove that the solution, if exists, of the equation stays the same sign as the initial 
value.  Section 3 is devoted to the derivation of the viscous Burgers equation for the mean correction function under the 
assumption that the density process of the Girsanov transform for the SDE possesses the path independent property. We 
also give some discussions of our notion of the path-independence and its link to terminologies in economics and finance. 
In Section 4, we develop the Galerkin type approximation for the derived viscous Burgers equation. 

\section{Preliminary}

We start with a brief account of Girsanov theorem. Given a complete probability space $(\Omega,\mathcal{F},\mathbb{P})$ 
with a usual  filtration $\{\mathcal{F}_t\}_{t\in[0,\infty)}$, let $b:[0,\infty)\times\mathbb{R}\to\mathbb{R}$
and $\sigma:[0,\infty)\times\mathbb{R}\to\mathbb{R}$ be
measurable functions. Let $\mathbb{E}$ denote the expectation with respect to the probability measure $\mathbb{P}$. We 
consider the following SDE 
\begin{equation} \label{SDE}
dX_t=b(t,X_t)dt+\sigma(t,X_t)dB_t, \quad t>0
\end{equation}
where $B_t$ is a standard Brownian motion.
It is well-known, see e.g. \cite[Theorem IV.3.1]{IkWa}, that when  $b$ and $\sigma$  satisfy a linear
growth and local Lipschitz condition with respect to the second variable,
there exists a unique solution to Equation (\ref{SDE}) with any given initial
data $X_0$ and the solution $X=(X_t)_{t\in[0,\infty)}$ is a real-valued continuous Markov process.  

The celebrated Girsanov theorem provides a very powerful
probabilistic tool to solve Equation (\ref{SDE}) under the name of
the {\it Girsanov transformation or the transformation of the
drift}. Let $\gamma:[0,\infty)\times\mathbb{R}\to\mathbb{R}$
satisfy the following Novikov condition
$$\mathbb{E}\left[\exp\left(\frac{1}{2}\int_0^t\gamma(s,X_s)^2ds\right)\right]
<\infty, \quad \forall t>0.$$ Then, by Girsanov theorem (cf e.g.
Theorem IV 4.1 of \cite{IkWa}),
$$\exp\left(\int^t_0\gamma(s,X_s)dB_s-\frac{1}{2}\int^t_0
\gamma(s,X_s)^2ds\right),\quad t\in[0,\infty)$$ is an
$\{\mathcal{F}_t\}$-martingale. Furthermore, for $t\ge0$, we define
$$\mathbb{Q}_t:=\exp\left(\int^t_0\gamma(s,X_s)dB_s-\frac{1}{2}
\int^t_0\gamma(s,X_s)^2ds\right)\cdot \mathbb{P}$$ or equivalently, in terms
of the Radon-Nikodym derivative
$$\frac{d\mathbb{Q}_t}{d\mathbb{P}}=\exp\left(\int^t_0\gamma(s,X_s), dB_s
-\frac{1}{2}\int^t_0\gamma(s,X_s)^2ds\right).$$ Then, for any
$T>0$,
$$\tilde{B}_t:=B_t-\int^t_0\gamma(s,X_s)ds,\quad 0\le t\le T$$
is an $\{\mathcal{F}_t\}$-Brownian motion under the probability
$\mathbb{Q}_T$. Moreover, $X_t$ satisfies

\begin{equation} \label{SDE-a}
dX_t=[b(t,X_t)+\sigma(t,X_t)\gamma(t,X_t)]dt
+\sigma(t,X_t)d\tilde{B}_t, \quad t>0.
\end{equation}
One can then discuss comprehensively the {\it existence and
uniqueness as well as the structure of solutions} to the initial
value problem for Equation (\ref{SDE}) by appealing the above
argument with suitable choice of $\gamma$. A remarkable choice of $\gamma$ is to 
vanish the drift coefficient. To be more precise, assume that $\sigma(t,x)\neq0$ for all 
$(t,x)\in[0,\infty)\times\mathbb{R}$. We then take 
\begin{equation}
\gamma(t,x):=-\frac{b(t,x)}{\sigma(t,x)}, \quad (t,x)\in[0,\infty)\times\mathbb{R}. 
\end{equation} 
Then, under the assumption of Novikov condition for this specified $\gamma$
$$\mathbb{E}\left[\exp\left(\frac{1}{2}\int_0^t \left[\frac{b(s,X_s)}{\sigma(s,X_s)}\right]^2ds\right)\right]
<\infty, \quad \forall t>0.$$
Equation ({\ref{SDE-a}}) becomes 
$$dX_t=\sigma(t,X_t)d\tilde{B}_t, \quad t\geq 0$$
indicating that $(X_t)_{t\in[0,T]}$ then becomes a local martingale on the probability set-up 
$(\Omega,\mathcal{F},\mathbb{Q}_T;\{\mathcal{F}_t\}_{t\in[0,T]})$.

In the present paper, we are concerned with the following SDE of mean-reversion type  
\begin{equation}\label{main equation}
	dR_t= (\theta +\sigma \alpha(R_t, t))R_t dt +\sigma R_t dB_t,\quad t>0
\end{equation}
where $\theta\in \mathbb{R}$ and $\sigma>0$ are constants, and the function $\alpha\in C^{2,1}(\mathbb{R}\times[0,\infty)) $. 
Here and in the sequel, $C^{2,1}(\mathbb{R}\times[0,\infty))$ stands for the space of functions $f: \mathbb{R}\times[0,\infty)\to \mathbb{R} $ such that both $\frac{\partial f^2}{\partial x^2}(x,t)$ and $\frac{\partial f}{\partial t}(x,t)$ are continuous.

From the above general discussion, we know that under the linear growth and the Lipschitz conditions on both drift and diffusion coefficients, there is a unique solution to \eqref{main equation} with a given initial value $R_0=r_0\in\mathbb{R}$. So we assume that there exists a constant $c>0$ such that
$$| \alpha (x,t)x|\leq  c(1+|x|)$$
and 
$$|x\alpha (x,t)-y\alpha (y,t)|\leq c|x-y|$$
for all $x,y\in \mathbb{R}$ and $t\in[0,\infty)$.

The next result will be used in the rest of the paper. The result is also interesting in itself.
\begin{Proposition}\label{positive solution}
  Let $(R_t,t\geq 0)$ be the solution of \eqref{main equation} with an initial value $R_0=r_0\in \mathbb{R}\setminus\{0\}$. If 
  \begin{equation}\label{novikov} 
  \mathbb{E} \left[ \exp \left (\frac{1}{2}\int_0^t \left ( \frac{\theta}{\sigma}+ \alpha (R_s,s) \right)^2 ds \right   ) \right]<\infty, \quad \forall \, t>0,
   \end{equation}
  then 
  $$R_t \,r_0>0, \quad \text{for all } \, t\geq 0.$$
 That is, the solution $(R_t,t\geq 0)$ keeps the same sign as the initial value $r_0$.
\end{Proposition}

\proof

Set, for $t\geq 0$,
$$\mathbb{Q}_t:=\exp\left ( -\int_0^t \left(\frac{\theta}{\sigma}+ \alpha (R_s,s) \right) dB_s-\frac{1}{2}\int_0^t\left( \frac{\theta}{\sigma}+ \alpha (R_s,s) \right)^2 ds   \right ) \mathbb{P}$$
and 
$$\tilde {B_t}:=B_t+\int_0^t\left(\frac{\theta}{\sigma}+ \alpha (R_s,s)\right) ds.$$
Then by Girsanov Theorem, for any $T>0$, $(\tilde {B_t}, 0\leq t\leq T)$ is a Brownian motion under $\mathbb{Q}_T$ and the \eqref{main equation} becomes
\begin{equation}\label{main equation 2}
dR_t=\sigma R_t d\tilde {B_t}.
\end{equation}
Clearly, Equation \eqref{main equation 2} has the explicit solution 
$$R_t=r_0 \exp\left(\tilde {B_t}-\frac{1}{2}\sigma ^2 t\right), \quad \, \forall t\geq 0$$
which justifies our claim. 
\qed

In the rest of the paper, we assume the condition \eqref{novikov} holds. We also assume that the initial value $r_0>0$. Hence, the process  $(R_t,t\geq 0)$ takes positive values only.

\section{The link of $\alpha$ with viscous Burgers equaiton} \label{Sec_RP}

This section is devoted to derive a non-linear partial differential equation of Burgers type for the mean correction term $\alpha$ in Equation \eqref{main equation}.

\begin{Definition}[{\bf Path independence of Girsanov transform density}]
Let the process $(R_t, t\ge0)$ be determined by Equation \eqref{main equation}.
We say that the Girsanov transform density (i.e. Radon-Nikodym derivative)
associated with the mean-reversion drift
$$ \frac{d\mathbb{Q}_t}{d\mathbb{P}} =\exp \left(- \frac{1}{2} \int_0^t\left(\frac{\theta+\sigma\alpha(R_s,s)}{\sigma}\right)^2ds
     - \int^t_0\frac{\theta+\sigma\alpha(R_s,s)}{\sigma}dB_s \right)$$
has path independent property if there exists  $F\in C^{2,1}(\mathbb{R}\times[0,\infty))$ such that
\begin{eqnarray*}
F(R_t,t)&=&F(R_0,0)- \frac{1}{2} \int_0^t\left(\frac{\theta+\sigma\alpha(R_s,s)}{\sigma}\right)^2ds \\
  && \quad  - \int^t_0\frac{\theta+\sigma\alpha(R_s,s)}{\sigma}dB_s , \quad \forall t>0\, .
\end{eqnarray*}
This is equivalent to say that the exponent of the Radon-Nikodym derivative is path independent, i.e.
\begin{equation}\label{RN}
 \ln\left(\frac{d\mathbb{Q}_t}{d\mathbb{P}}\right) =F(R_t,t)-F(R_0,0),\quad \text{for all }\, t>0.
\end{equation}
\end{Definition}

The concept of {\it path independence} of Girsanov transform density has its root in mathematical economics. In a market with underlying stock price dynamics described by SDEs of mean reversion type, the {\it market efficiency} is indeed characterised by the path independent property of certain utility function which can be expressed in terms of Girsanov transform density, see c.f e.g. \cite{SHAC, Stein}. Market efficiency is sometime linked with the terminology {\it market equilibrium} in certain literature.

Let us explicate this point further.  A conventional kind of equilibrium market can be characterized by a value function $V:[0,\infty)\times \mathbb{R}\to \mathbb{R}$ of a representative agent (see e.g., \cite{HL1988,D19981,D19982, DR2003}).  Given the probability measure $P$ as an objective probability in the market model, one can interpret our $R_t$ as the wealth of the representative agent in a single stock market. Assuming that the representative agent has certain utility function $U$, depending on time $t$ and stock price process $(R_t, t\geq 0)$, and  maximizes his expected total utility and that the value function $V$ is differentiable in the first variable and defined as the expectation of total utility, Cox and Leland in \cite{CL1982} show that the path
independence is necessary for expected utility maximization. By path independence, they mean that the value of a portfolio at certain time $t>0$
will depend only on the asset prices at time $t$, not on the path followed by the asset in reaching that price. Namely, the utility $U$ depends on the state price $R_t$ and time $t$ that is, the function $U$ is of the form $U(t, R_t)$. Further more, it was shown e.g. in \cite{SHAC} that there exists a risk neutral probability measure $\mathbb{Q}$ which is absolutely continuous with respect to $\mathbb{P}$ and that the Radon-Nikodym derivative $\frac{d\mathbb{Q}}{d\mathbb{P}}$ gives the state-price density. Combining the above $U(t,R_t)$, therefore, the Radon-Nikodym derivative is exactly in the form of \eqref{RN}.

We are now in the position to present our main result of this section.
\begin{Theorem} \label{Theor_alpha}

Let $(R_t,t\ge0)$ be the solution of \eqref{main equation} with an initial value $R_0=r_0>0$. If the Girsanov transform density of \eqref{main equation} has the path independent property, then the mean correction function $\alpha$ satisfies the time-reverse
viscous Burgers equation
\begin{equation} \label{eq_alpha_t1}
    \frac {\partial}{\partial t} \alpha (x,t) = - \frac {1}{2}   \sigma ^2 \frac {\partial ^2 }{\partial x^2} \alpha (x,t) - \sigma \alpha (x,t) \frac{\partial}{\partial x}\alpha (x,t)
\end{equation}
for $(x,t)\in\mathbb{R}\times[0,\infty)$.
\end{Theorem}

\medskip

\proof

From Proposition \ref{positive solution}, we know $R_t>0$ for all $t>0$ since $r_0>0$. Then we define a process $(X_t,t\geq 0)$ by
\begin{equation} \label{eq_transformX}
    X_t:= \ln R_t - (\theta - \frac{1}{2}\sigma ^2) t.
\end{equation}
 Treating $R_t$ as an integrity variable and using It\^{o} formula to (\ref{eq_transformX}), together with Equation \eqref{main equation}, we get
\begin{equation} \label{eq_dX_t}
\begin{split}
    dX_t &=\frac{\partial X_t}{\partial t} dt +\frac{\partial X_t}{\partial R_t} dR_t + \frac{1}{2} \frac{\partial ^2 X_t}{\partial R_t^2} (dR_t)^2 \\
    &= -(\theta -\frac{1}{2} \sigma ^2) dt + \frac{1}{R_t} ((\theta + \sigma \alpha (t,R_t)) R_t dt +\sigma R_t dB_t)\\
	&\quad + \frac{1}{2} (-1) \frac{1}{R_t^2} \sigma ^2 R_t^2 dt \\
    &= -(\theta -\frac{1}{2} \sigma ^2) dt +(\theta + \sigma \alpha (t,R_t)) R_t dt +\sigma dB_t - \frac{1}{2} \sigma ^2 dt \\
    &= \sigma \alpha(t, R_t) dt + \sigma dB_t.
\end{split}
\end{equation}
By Girsanov theorem, we know that the change of measure is characterized by the Radon-Nikodym  derivative
\[
    \frac{d\mathbb{Q}_t}{d\mathbb{P}} =\exp \left(- \frac{1}{2} \int_0^t \alpha ^2 (X_s,s)ds - \int^t_0\alpha(X_s,s)dB_s \right).
\]
Under the assumption that the Girsanov transform density of \eqref{main equation} has the path independent property, we get
$$\frac{1}{2}  \int_0^t \alpha ^2 (X_s,s)ds + \int^t_0\alpha(X_s,s)dB_s $$
is also path independent. Set
\begin{equation} \label{eq_BT}
    Z(X_t,t) :=\frac{1}{2}  \int_0^t \alpha ^2 (X_s,s)ds + \int^t_0\alpha(X_s,s)dB_s
\end{equation}
then we can write
\[
    \frac{d\mathbb{Q}_t}{d\mathbb{P}} =\exp (- Z(X_t,t)).
\]
From Equation (\ref{eq_dX_t}), we have
\begin{equation} \label{eq_dB_t}
    dB_t = \frac{dX_t - \sigma \alpha (X_t,t)dt}{\sigma},
\end{equation}
further substituting Equation (\ref{eq_dB_t}) into the integrand of (\ref{eq_BT}) in place of $dB_t$ then yields
\begin{equation} \label{eq_Z(t,X_t)}
\begin{split}
    Z(X_t,t) &= \int_0^t \Big (\frac{1}{2}\alpha ^2 (X_s,s)ds + \alpha (X_s,s) \frac{dX_s - \sigma \alpha (X_s,s)ds}{\sigma } \Big )\\
	&= \int_0^t \Big (-\frac{1}{2}\alpha ^2 (X_s,s)ds + \frac{\alpha  (X_s,s)}{\sigma} dX_s \Big ).
\end{split}
\end{equation}
Differential formulation of \eqref{eq_Z(t,X_t)} is written as 
\begin{equation} \label{eq_dZ_t1}
    dZ(X_t,t) = -\frac{1}{2}\alpha ^2 (X_t,t)dt + \frac{\alpha  (X_t,t)}{\sigma} dX_t.
\end{equation}
On the other hand, treating $X_t$ as an integrity variable and using It\^{o} formula for the composition $ Z(X_t,t)$, we derive that
\begin{equation} \label{eq_dZ_t2}
    dZ(X_t,t) = \frac{\partial}{\partial t}Z(X_t,t)dt + \frac{\partial}{\partial X_t}Z(X_t,t)dX_t + \frac{1}{2}\sigma ^2 \frac{\partial ^2}{\partial X_t^2}Z(X_t,t)dt.
\end{equation}
Equating the right hand side of (\ref{eq_dZ_t1}) to that of  (\ref{eq_dZ_t2}), we conclude
\begin{align} \label{eq_equaldZ}
	 & - \frac{1}{2} \alpha ^2(X_t,t)dt + \frac{\alpha (X_t,t)}{\sigma }dX_t \nonumber\\
	=& \left ( \frac{\partial}{\partial t}Z(X_t,t)+ \frac{1}{2}\sigma ^2 \frac{\partial ^2}{\partial X_t^2}Z(X_t,t)\right )dt + \frac{\partial}{\partial X_t}Z(X_t,t)dX_t.
\end{align}
Comparing the coefficients of $dt$ and $dX_t$ in (\ref{eq_equaldZ}) respectively, we get
\begin{equation} \label{eq_Z_x}
    \frac{\partial}{\partial X_t}Z(X_t,t) = \frac {\alpha (X_t,t)}{\sigma}
\end{equation}
and
\begin{equation} \label{eq_Z_t}
    \frac{\partial}{\partial t}Z(X_t,t) + \frac{1}{2}\sigma ^2 \frac{\partial ^2}{\partial X_t^2}Z(X_t,t) = -\frac{1}{2}\alpha ^2 (X_t,t).
\end{equation}
We are aiming to find an equation for $\alpha$, so we try to eliminate $Z$ by manipulating Equations (\ref{eq_Z_x}) and (\ref{eq_Z_t}). Differentiating Equation (\ref{eq_Z_x}) again with respect to the variable $X_t$ we get
\begin{equation} \label{eq_Z_xx}
    \frac{\partial ^2}{\partial X_t^2}Z(X_t,t) = \frac {1}{\sigma}\frac{\partial}{\partial X_t}\alpha (X_t,t)
\end{equation}
then substituting Equation (\ref{eq_Z_xx}) into Equation (\ref{eq_Z_t}), we have
\begin{equation} \label{eq_Z_t2}
    \frac{\partial}{\partial t}Z(X_t,t) = - \frac {1}{2} \sigma \frac{\partial}{\partial X_t}\alpha (X_t,t) - \frac {1}{2}\alpha ^2 (X_t,t)\, .
\end{equation}
Now, to eliminate $Z$, we can differentiate Equation (\ref{eq_Z_t2}) with respect to $x$ and (\ref{eq_Z_x}) with respect to $t$, respectively,
\[
    \frac{\partial ^2}{\partial t \partial X_t}Z(X_t,t) = - \frac {1}{2} \sigma \frac{\partial ^2}{\partial X_t^2}\alpha (X_t,t) - \alpha (X_t,t) \frac{\partial}{\partial X_t}\alpha (X_t,t)
\]
and
\[
    \frac{\partial ^2}{\partial X_t \partial t}Z(X_t,t) = \frac{1}{\sigma}\frac {\partial}{\partial t}\alpha (X_t,t).
\]
Furthermore, equating the above two equations, we get
\begin{equation*}
    \frac {\partial}{\partial t} \alpha (X_t,t) = - \frac {1}{2}   \sigma ^2 \frac {\partial ^2 }{\partial X_t^2} \alpha (X_t,t) - \sigma \alpha (X_t,t) \frac{\partial}{\partial X_t}\alpha (X_t,t).
\end{equation*}

Since $\sigma>0$, the solution $(R_t,t\geq 0)$ of \eqref{main equation} is non-degenerate.  By the definition \eqref{eq_transformX}, the process $(X_t, t\geq 0)$ is fully supported on $\mathbb{R}$. Hence we obtain
\begin{equation*}
    \frac {\partial}{\partial t} \alpha (x,t) = - \frac {1}{2}   \sigma ^2 \frac {\partial ^2 }{\partial x^2} \alpha (x,t) - \sigma \alpha (x,t) \frac{\partial}{\partial x}\alpha (x,t).
\end{equation*}
This shows that the  mean correction function $\alpha$ in \eqref{main equation} satisfies the time-reversal
viscous Burgers equation. We are done. \qed

\begin{Remark}
The (viscous) Burgers equation arises in connection with the behavior of the risk premium of the market portfolio of risky assets in the classical  Black-Scholes efficient stock market models can be found in Bick \cite{Bick} and He-Leland \cite{HeLeland}. Our analysis in this section is inspired by Hodges-Carverhill \cite{SHAC} and Hodges-Selby \cite{HodgeSelby}, which investigate the behavior of a single asset price $S_t$ in an equilibrium market with price dynamics follows the following type stochastic differential equation
\[
    \frac{dS}{S}=[r+ \sigma \alpha(S,t)]dt + \sigma dz
\]
where $r$ is a constant risk-free interest rate, $\sigma$ is a constant volatility parameter, $z$ stands for a Brownian motion, and $\alpha(S,t)$
is an adapted stochastic process standing for the risk price. Hodges \textit{et al.} in \cite{SHAC,HodgeSelby} claim that in the equilibrium market, the risk premium  $\alpha(S,t)$ must satisfy the (viscous) Burgers equation. Let us also mention an interesting articles \cite{Stein} for the relevant analytic
studies for the Black-Scholes' equilibrium market model with the above type stochastic differential equation. 
\end{Remark} 

\begin{Remark} 
The characterization of path independent property of the density processes of Girsanov transformation for SDEs with more general 
coefficients and for SDEs in multi-dimensional spaces as well as for SDEs on differential manifolds can be found in \cite{WuY,TWWY}.  
\end{Remark}

\section{An approximation scheme to the mean correction term $\alpha$}

Tracing back to 1950 and 1951, Hopf and Cole derived independently an analytic solution to the Burgers equation by using a transformation, nowadays called {\it Hopf-Cole transformation}, which reduces the Burgers equation (which is a nonlinear partial differential equation) to a heat diffusion equation (which is linear). Namely, the (viscous) Burgers equation can be solved in closed form in terms of the initial data by utilising the Hopf-Cole substitution. However, to establish an applicable, numerical solution to the initial value problem for the Burgers equation is a very interesting problem and it remains as a great challenge as long as applying the Burgers equation to model various practical problems. Thereafter, there is an increasing interest to solve Burgers equation numerically. Majda and Timofeyev \cite{MT,MT1} introduced a very remarkable method which provides a suitable approximation of the inviscid Burgers equation. Their approximation involves Galerkin projection on the Fourier modes involved. Our analysis in this section is inspired by \cite{MT,MT1}  with resulting in a numerical approximation to our formerly derived viscous Burgers equation.

We recall our viscous Burgers equation is given as follows
\begin{equation} \label{eq_Burger4}
    \frac {\partial}{\partial t} \alpha (x,t) = - \frac {1}{2}   \sigma ^2 \frac {\partial ^2 }{\partial x^2} \alpha (x,t) - \sigma \alpha (x,t) \frac{\partial}{\partial x}\alpha (x,t),\, (x,t)\in\mathbb{R}\times[0,\infty). 
\end{equation}
and rewrite this equation as
\begin{equation} \label{eq_alpha_t2}
	\frac {\partial}{\partial t} \alpha (x,t) = - \frac {1}{2}   \sigma ^2 \frac {\partial ^2 }{\partial x^2} \alpha (x,t) -\frac{\sigma}{2} \frac{\partial}{\partial x} \big[\alpha (x,t) \big]^2.
\end{equation}
Using the Fourier transformation, one can convert this equation to an ordinary differential equation. Define the Fourier transformation and the inverse Fourier transformation of $\alpha$
as
\[
	\hat{\alpha}(k,t):=\frac{1}{\sqrt{2\pi}}\int_{-\infty}^{\infty}e^{-ikx}\alpha(x,t)dx
\]
and
\[
	\alpha (x,t) := \frac{1}{\sqrt{2\pi}}\int_{-\infty}^{\infty}e^{ikx}\hat{\alpha}(k,t)dk
\]
provided the both integral are well-defined. We have the following two properties:
\begin{align*}
    &(i) \quad \frac{\partial}{\partial k}\hat{\alpha}(k,t)= ik\hat{\alpha}(k,t) \\
    &(ii) \quad \frac{\partial^2}{\partial k^2}\hat{\alpha}(k,t)= (ik)^2\hat{\alpha}(k,t)=-k^2\hat{\alpha}(k,t).
\end{align*}
After the Fourier transformation, (\ref{eq_alpha_t2}) becomes
\begin{equation} \label{eq_hatalpha_t1}
	\frac{\partial}{\partial t}\hat{\alpha}(k,t)=-\frac{1}{2}\sigma^2(ik)^2\hat{\alpha}(k,t)-\frac{\sigma}{2}(ik)\widehat{\alpha^2}(k,t)\, .
\end{equation}
Furthermore, discretising the Fourier transformation into the Fourier series and truncating the Fourier series, we get an approximation to solution $\hat{\alpha}(k,t)$ of the ordinary differential equation (\ref{eq_hatalpha_t1}), which in turn provides an approximation to the solution
$\alpha(k,t)$ of the viscous Burgers equation (\ref{eq_Burger4}). Define now the Galerkin truncation for arbitrarily fixed natural number $N\in\mathbb{N}$ as follows
\[
	\alpha_N(x,t):=(P_N\alpha)(x,t) := \sum ^N_{k=-N} \hat{\alpha}(k,t)e^{ikx}.
\]
Then it is clear that
\[
	\frac{\partial ^2}{\partial x^2} \alpha_N(x,t)=(P_N\frac{\partial^2}{\partial x^2}\alpha)(x,t)=\sum^N_{k=-N} (ik)^2\hat{\alpha} (k,t)e^{ikx}
\]
and
\begin{align*}
    \frac{\partial}{\partial x} \left[ \alpha ^2_N(x,t)\right] &= P_N \left[\frac{\partial}{\partial x}(\alpha ^2(x,t))\right]  \\
    &= \sum^N_{k=-N} (ik) \hat{\alpha}(k,t) \hat{\alpha}(k,t)e^{ikx}\\
    &= \sum^N_{k=-N} (ik) \sum^N_{p=-N}\hat{\alpha}(p,t) \hat{\alpha}(k-p,t)e^{ikx}.
\end{align*}
Consequently, Equation (\ref{eq_hatalpha_t1}) becomes
\begin{align*}
	\frac{\partial}{\partial t}(\sum^N_{k=-N} \hat{\alpha}(k,t)e^{ikx})=& - \frac{\sigma ^2}{2}\sum^N_{k=-N} (ik)^2\hat{\alpha} (k,t)e^{ikx} \\
	& -\frac{\sigma}{2}\sum^N_{k=-N} (ik) \sum^N_{p=-N}\hat{\alpha}(p,t) \hat{\alpha}(k-p,t)e^{ikx}\, .
\end{align*}
Simplifying the above equation, we get
\begin{align*}
	\sum^N_{k=-N} \frac{\partial}{\partial t}\hat{\alpha}(k,t)e^{ikx} &= \frac{\sigma ^2}{2}\sum^N_{k=-N} k^2\hat{\alpha} (k,t)e^{ikx} \\
	&\quad -\frac{\sigma}{2}\sum^N_{k=-N} (ik) \sum^N_{p=-N}\hat{\alpha}(p,t) \hat{\alpha}(k-p,t)e^{ikx}\\
	&= \sum^N_{k=-N}\left( \frac{\sigma ^2}{2}k^2 \hat{\alpha} (k,t)-\frac{\sigma ik}{2}\sum^N_{p=-N}\hat{\alpha}(p,t) \hat{\alpha}(k-p,t)\right)
            e^{ikx}.
\end{align*}
Fixing k and comparing the coefficients on the both sides, we get
\begin{equation} \label{eq_hatalpha_t2}
	\frac{\partial}{\partial t} \hat{\alpha}(k,t)= \frac{\sigma ^2}{2}k^2 \hat{\alpha} (k,t)-\frac{\sigma ik}{2}\sum^N_{p=-N}\hat{\alpha}(p,t)
     \hat{\alpha}(k-p,t)
\end{equation}
for $k=-N,-N+1,...,N-1,N$. It turns out that the partial differential equation (\ref{eq_Burger4}) has been reduced to a system of $2N+1$ ordinary differential equations with the help of truncated (discretised) Fourier transformation. Then we may rewrite Equation (\ref{eq_hatalpha_t2}) in the conventional manner

\begin{equation} \label{eq_dalpha_t1}
	\frac{d}{dt}\hat{\alpha}(k,t)=\frac{\sigma ^2}{2}k^2 \hat{\alpha} (k,t)-\frac{\sigma ik}{2}\sum^N_{p=-N}\hat{\alpha}(p,t) \hat{\alpha}(k-p,t).
\end{equation}

Next, we shall try to find a numerical solution to the Equation (\ref{eq_dalpha_t1}). For any natural number $n\in\mathbb{N}$, we split the time interval $[0,T]$ into $n$ equal-sized sub-intervals, i.e.
\[
	0=t_0<t_1<t_2<\cdots <t_{n-1}<t_n=T
\]
and denote the step-size $\delta:=\frac{T-0}{n}$ as a fixed parameter. Replacing $\frac{d}{dt}\hat{\alpha}(k,t)$ by the forward numerical differentiation $\frac{\hat{\alpha}(k,t_j)-\hat{\alpha}(k,t_{j-1})}{\delta}$, where $j=1,2,\cdots,n$, and putting it into Equation (\ref{eq_dalpha_t1}), we get
\[
	\frac{\hat{\alpha}(k,t_j)-\hat{\alpha}(k,t_{j-1})}{\delta}=\frac{\sigma ^2}{2}k^2 \hat{\alpha} (k,t_{j-1})-\frac{\sigma ik}{2}\sum^N_{p=-N}
\hat{\alpha}(p,t_{j-1}) \hat{\alpha}(k-p,t_{j-1})\, .
\]
We now rearrange this equation as follows
\begin{equation} \label{eq_hatalpha}
	\hat{\alpha}(k,t_j)=\left(\frac{\delta\sigma^2}{2}k^2+1\right)\hat{\alpha}(k,t_{j-1})-\frac{\delta\sigma ik}{2}\sum^N_{p=-N}\hat{\alpha}(p,t_{j-1})\hat{\alpha}(k-p,t_{j-1})\, , 
\end{equation}
for all $k=-N,-N+1,...,N-1,N$. This is a recurrence formula on $t_0, t_1,\ldots, t_n$, which is a Galerkin truncation of the equation (\ref{eq_hatalpha_t1}). Namely, the set $\{\hat{\alpha}(k,t_0), \hat{\alpha}(k,t_1), \ldots, \hat{\alpha}(k,t_n)\}$ forms a numerical solution to the equation (\ref{eq_dalpha_t1}) in the sense that $\hat{\alpha}(k,t_n)$ converges stably to the Fourier transformation of the analytic closed form solution of the initial value problem for the equation (\ref{eq_Burger4}) as the step-size $\delta\to0$ or equivalently as $n\to\infty$. The
mathematical justification of this argument is rather routine in the topic of numerical solutions to differential equations (cf. e.g. \cite{KlPl}).

Finally we come to the stage to get the initial value $\hat{\alpha}(k,t_0)=\hat{\alpha}(k,0)$,  for each $k=-N,-N+1,...,N-1,N$. With the given initial data of real-valued  $\alpha (x,0)$, we have
\[
	\alpha(x,0)=\alpha _N(x,0)=\sum^N_{m=-N}\hat{\alpha}(m,0)e^{imx}.
\]
Now for any fixed $-N\le k\le N$, multiplying $e^{-ikx}$ to the above identity and taking integral of $x$ over $[-\pi,\pi]$, we get 
$$\int^{\pi}_{-\pi}\alpha(x,0)e^{-ikx}dx= \sum^N_{m=-N}\hat{\alpha}(m,0)\int_{-\pi}^\pi e^{imx}e^{-ikx}dx=2\pi\hat{\alpha}(k,0)$$
since 
$$\int_{-\pi}^\pi e^{ikx}e^{-ikx}dx=2\pi\quad \mbox{and}\quad \int_{-\pi}^\pi e^{imx}e^{-ikx}dx=0\quad \mbox{for}\quad m\neq k.$$
Hence, we obtain 
\begin{equation} \label{eq_sol-alpha}
	\hat\alpha(k,0)=\frac{1}{2\pi}\int_{-\pi}^\pi\alpha(x,0)e^{-ikx}dx.
\end{equation}

To summarise, in this section, with Galerkin truncation of the (discretised) Fourier transformation of the viscous Burgers equation (\ref{eq_Burger4}), we derive a recurrence formula (\ref{eq_hatalpha}) with the initial value (\ref{eq_sol-alpha}) which is a numerical solution to the mean correction function $\alpha$.

\end{document}